\documentclass{webofc}
\usepackage[varg]{txfonts}   % Web of Conferences font
\usepackage{bm}
\newcommand{\snn}{\mbox{$\sqrt{s_{_{NN}}}$}\xspace}
\begin{document}
%
%\title{Global and local polarization of hyperons in heavy-ion collisions}
\title{Recent progress on global and local polarization of hyperons in heavy-ion collisions}

\author{\firstname{Takafumi} \lastname{Niida}\inst{1}\fnsep\thanks{\email{niida@bnl.gov 
    }} 
}

\institute{University of Tsukuba, 1-1-1 Tennoudai, Tsukuba, Ibaraki 305-8571, JAPAN
          }

\abstract{%
Since the observation of $\Lambda$ global polarization at RHIC, indicating the most vortical fluid ever observed so far, 
many experimental and theoretical progresses have been made in understanding of vorticity and polarization phenomena in heavy-ion collisions. 
But there still exist open questions which are currently being studied. In these proceedings, recent experimental results on 
global and local polarization of hyperons in heavy-ion collisions are reviewed.
}
\maketitle
\section{Introduction}\label{intro}
A partial conversion of an orbital angular momentum carried by two colliding nuclei into spin angular momentum of produced particles
is referred to as global polarization~\cite{Liang:2004ph,Voloshin:2004ha,Becattini:2007sr}. Observation of $\Lambda$ global polarization by STAR experiment~\cite{STAR:2017ckg} revealed 
the creation of most vortical fluid ever observed.
It will also help to constrain the lifetime of the initial magnetic field created in the collisions as the direction of the magnetic field coincides 
with the orbital angular momentum direction.
While the experimental results of average global polarization are well reproduced by theoretical models, most of which are based on local thermal equilibrium, 
there are open questions and some of differential measurements including local polarization are not yet fully understood.
Study of the polarization phenomena in heavy-ion collisions is relatively new but rapidly growing in this field, 
providing new information on the initial condition, medium properties, and the collision dynamics.
In these proceedings, recent results of hyperon polarization in heavy-ion collisions are reviewed.

\section{Global polarization}\label{sec-1}
Global polarization $P_H$, the polarization component along the initial orbital angular momentum carried by two colliding nuclei in heavy-ion collisions,
was first observed using $\Lambda$ hyperons in Au+Au collisions at $\sqrt{s_{NN}}$ = 7.7-200 GeV by STAR experiment~\cite{STAR:2017ckg,STAR:2018gyt}
where the polarization signal decreases with increasing the energy. There was also an attempt to measure it at the LHC energy but 
the result is consistent with zero as the expected signal is at the level of current precision.
Theoretical models predict that the polarization increases starting from \snn$\sim 2m_{N}$ and becomes a maximum around \snn = 3 GeV and 
the peak position could be different between $\Lambda$ and $\bar{\Lambda}$~\cite{Ayala:2021xrn}.

\begin{figure}
\begin{minipage}[t]{0.48\hsize}
	\centering
	\includegraphics[width=\linewidth,clip]{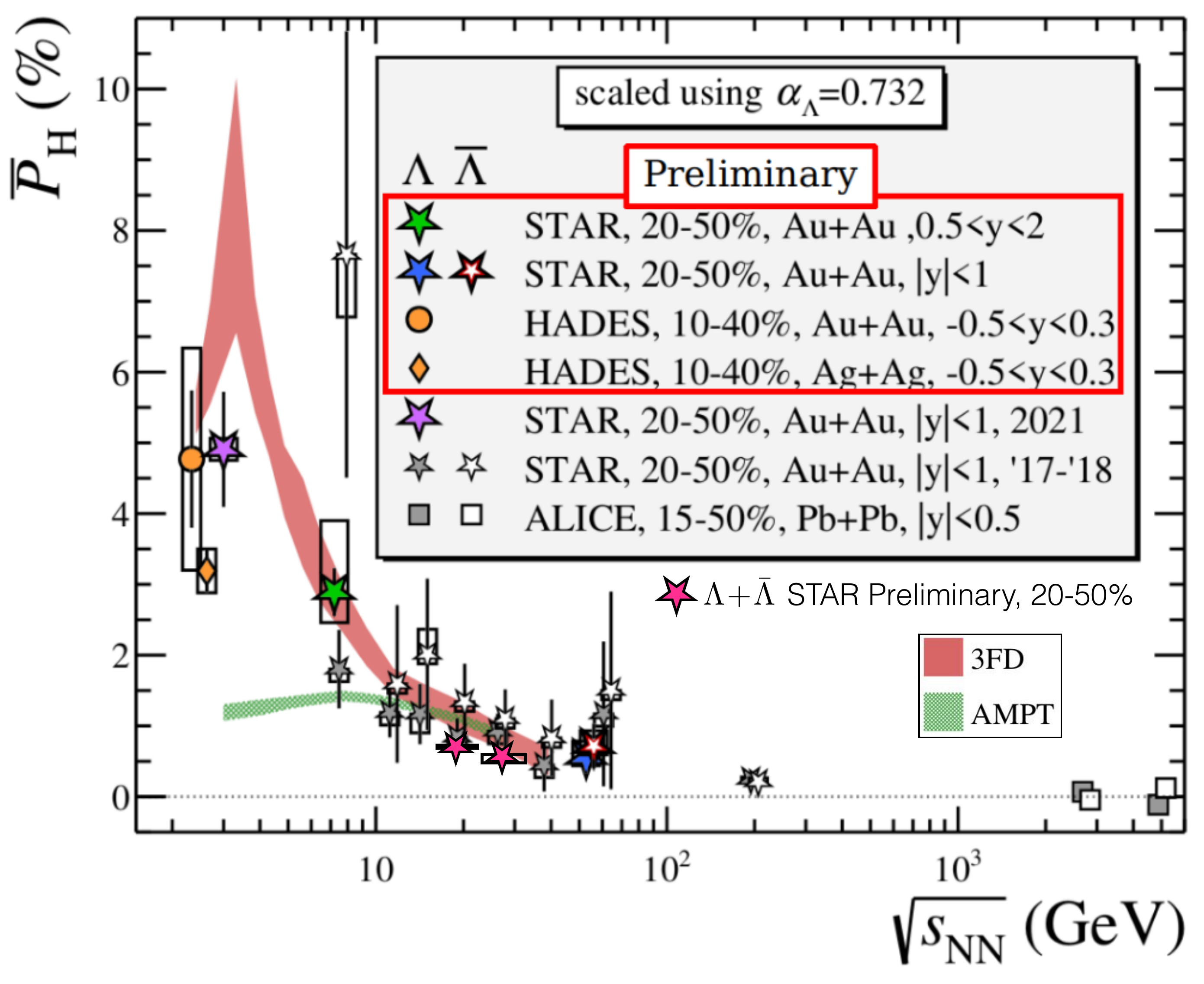}
\end{minipage}
\hspace{0.2mm}
\begin{minipage}[t]{0.50\hsize}
	\centering
	\includegraphics[width=\linewidth,clip]{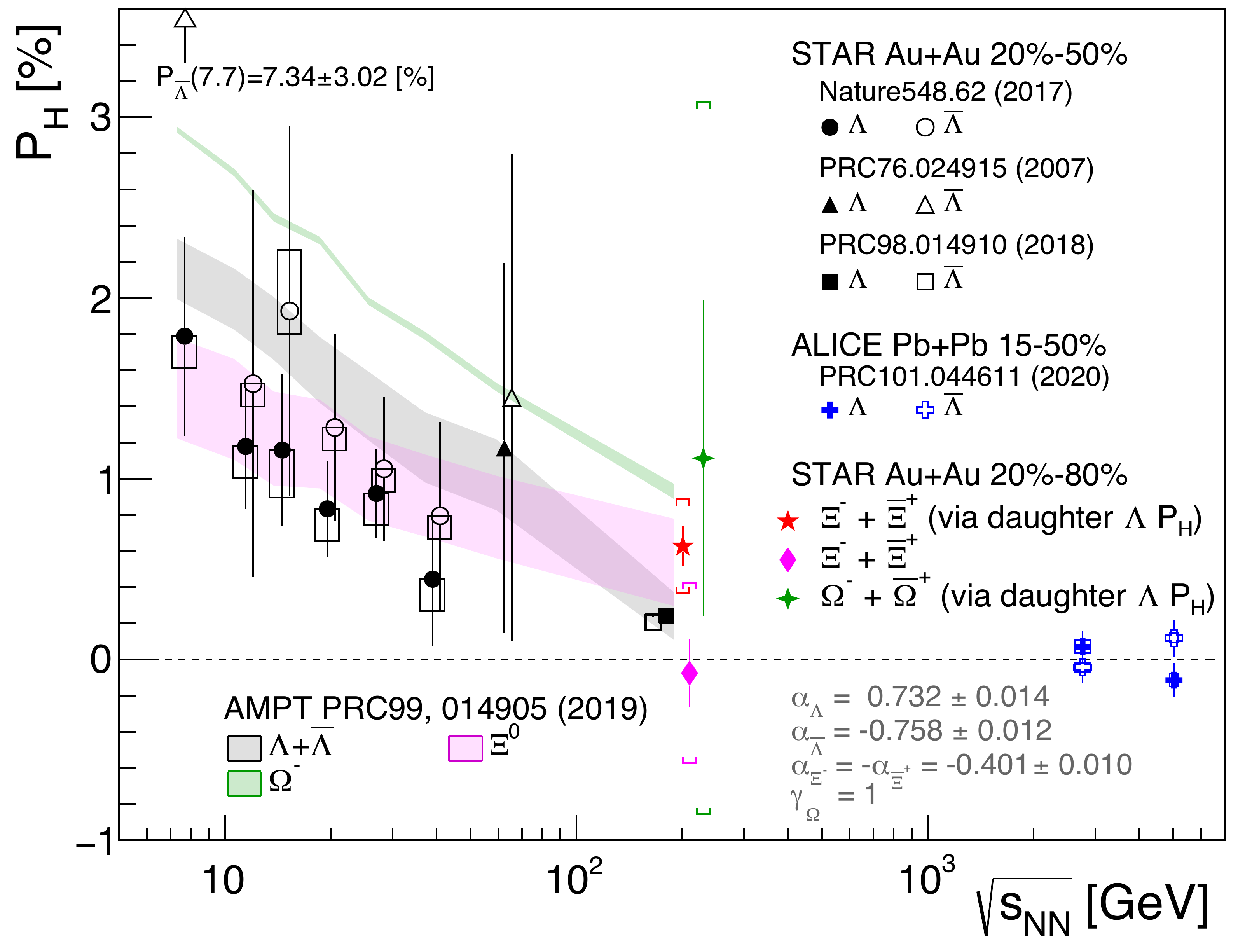}
\end{minipage}
\vspace{-0.2cm}
\caption{(Left) Global polarization of $\Lambda$ and $\bar{\Lambda}$ hyperons as a function of the center-of-mass energy 
from STAR, ALICE, and HADES experiments~\cite{qm2022:Joey}. 
(Right) global polarization of $\Xi$ and $\Omega$ hyperons compared to that of $\Lambda$ hyperons as a function of the collision energy~\cite{STAR:2020xbm}.}
\label{PH} 
\end{figure}

Recently, the measurement has been extended to lower energies by STAR fixed-target mode and HADES experiment~\cite{STAR:2021beb,HADES:2022enx}.
Figure~\ref{PH}(left) shows a compilation of $\Lambda$ global polarization as a function of the collision energy. New results at 3 GeV from STAR and
at 2.4 and 2.55 GeV from HADES still show an increasing trend of the global polarization,
consistent with the calculation from the hydrodynamic model based on three-fluid dynamics~\cite{Ivanov:2020udj}.
Note that the lower energy corresponds to the high baryon density region where the medium properties would be notably different. 
It should be also noted that the detector acceptance in rapidity and transverse momentum $p_T$ affects the energy dependence 
since it covers only mid-rapidity at higher energies while it does a nearly whole distribution of produced particles at lower energies.
Theoretical models predict differently in the rapidity dependence of global polarization 
although the current measurements do not show any significant dependence~\cite{STAR:2018gyt,STAR:2021beb,HADES:2022enx}.
Difference between $\Lambda$ and $\bar{\Lambda}$, as a measure of the initial magnetic field, is not significant with large uncertainties
which will be improved in the beam energy scan phase-II at RHIC-STAR.

To confirm the vorticity and global polarization picture, it is important to have the measurements for different particle species, 
preferably with different spins, which provides new insights into the polarization mechanism in heavy-ion collisions.
Recently, the measurements with multistrangeness, i.e. $\Xi$ (spin-1/2) and $\Omega$ (spin-3/2), have been performed 
by STAR experiment~\cite{STAR:2020xbm}. Unlike for $\Lambda$ hyperons ($\alpha_\Lambda=0.732$), the magnitude of decay parameter is small for $\Xi$ ($\alpha_\Xi=-0.401$) 
and is close to zero for $\Omega$ ($\alpha_\Omega=0.0157$) which makes the polarization measurement difficult.
One can use the measurement of daughter $\Lambda$ polarization to know its parent particle's polarization:
$P_{\rm \Lambda}^\ast = C_{\rm X\Lambda} P_{\rm X}^{\ast}$
where $P_{\rm X}^\ast$ and $P_{\rm \Lambda}^\ast$ are polarizations of parent particle X and its daughter $\Lambda$ in parent rest frame, 
and $C_{\rm X\Lambda}$ denotes polarization transfer factor. The factor $C_{\rm X\Lambda}$ is +0.944 for the decay of $\Xi^-\rightarrow \Lambda\pi^-$ 
and could be either +1 or -0.6 for the decay of $\Omega^-\rightarrow \Lambda K^-$ due to the uncertainty of decay parameter $\gamma_{\Omega}(\approx \pm 1)$.
Based on the vorticity picture, the sign of $\gamma_\Omega$ can be determined by the measurement of $\Omega$ global polarization.

Figure~\ref{PH}(right) shows global polarization of $\Xi$ and $\Omega$ hyperons comparing to $\Lambda$ global polarization 
as well as transport model calculations. The global polarization averaged over transverse momentum and rapidity 
for 20-80\% Au+Au collisions at \snn = 200 GeV is measured to be
$\langle P_{\Lambda}\rangle=0.24\pm0.03({\rm stat})\pm0.03({\rm stat})$\%, $\langle P_{\Xi}\rangle=0.47\pm0.10({\rm stat})\pm0.23({\rm stat})$\%,
and $\langle P_{\Omega}\rangle=1.11\pm0.87({\rm stat})\pm1.97({\rm stat})$\%.
Based on thermal model in the nonrelativistic limit~\cite{Becattini:2016gvu}, the polarization can be given by 
\begin{equation}
{\bm P}=\langle {\bm s} \rangle/s \approx (s+1){\bm \omega}/(3T),
\label{eq:P-spin}
\end{equation}
where $s$ is the spin of the particle, ${\bm \omega}$ is the local vorticity of the fluid, and $T$ is the temperature.
There seems hierarchy in the measured $\langle P_{\rm H}\rangle$ as expected from Eq.~\eqref{eq:P-spin} 
although the uncertainty is still large to make definitive conclusion.
The current measurement seems to favor the decay parameter $\gamma_\Omega$ to be $+1$.
More precise measurements in the future runs at RHIC and the LHC will help to shed light on the uncertainty of $\gamma_\Omega$
as well as for better understanding of the polarization dependence on particle species.

\section{Local polarization}
In general, local velocity would depend on many other things, such as density fluctuations, the effect of jets passing through the medium, and collective flow.
Thus vorticity could be created locally in heavy-ion collisions~\cite{Pang:2016igs,Betz:2007kg,Xia:2018tes}.
Vorticity along the beam direction, therefore the polarization, is expected to be induced by elliptic flow, i.e. stronger expansion (more particle emission) 
in the reaction plane direction than in the out-of-plane direction~\cite{Voloshin:2017kqp,Becattini:2017gcx}.
Such a longitudinal polarization was first observed in Au+Au collisions at \snn = 200 GeV by STAR experiment~\cite{STAR:2019erd}, and 
later in Pb+Pb collisions at \snn = 5.02 TeV by ALICE experiment~\cite{ALICE:2021pzu}.
There have been extensive theoretical studies to understand the data since the hydrodynamic and transport models fails to explain the sign of the observed signal,
although they reasonably describe the data of the global polarization. New term of shear tensor proposed in Refs.~\cite{Becattini:2021iol,Fu:2021pok} seems to be necessary 
to explain the data but further experimental inputs are needed for better understanding. 

\begin{figure}
\begin{minipage}[t]{0.49\hsize}
	\centering
	\includegraphics[width=\linewidth,clip]{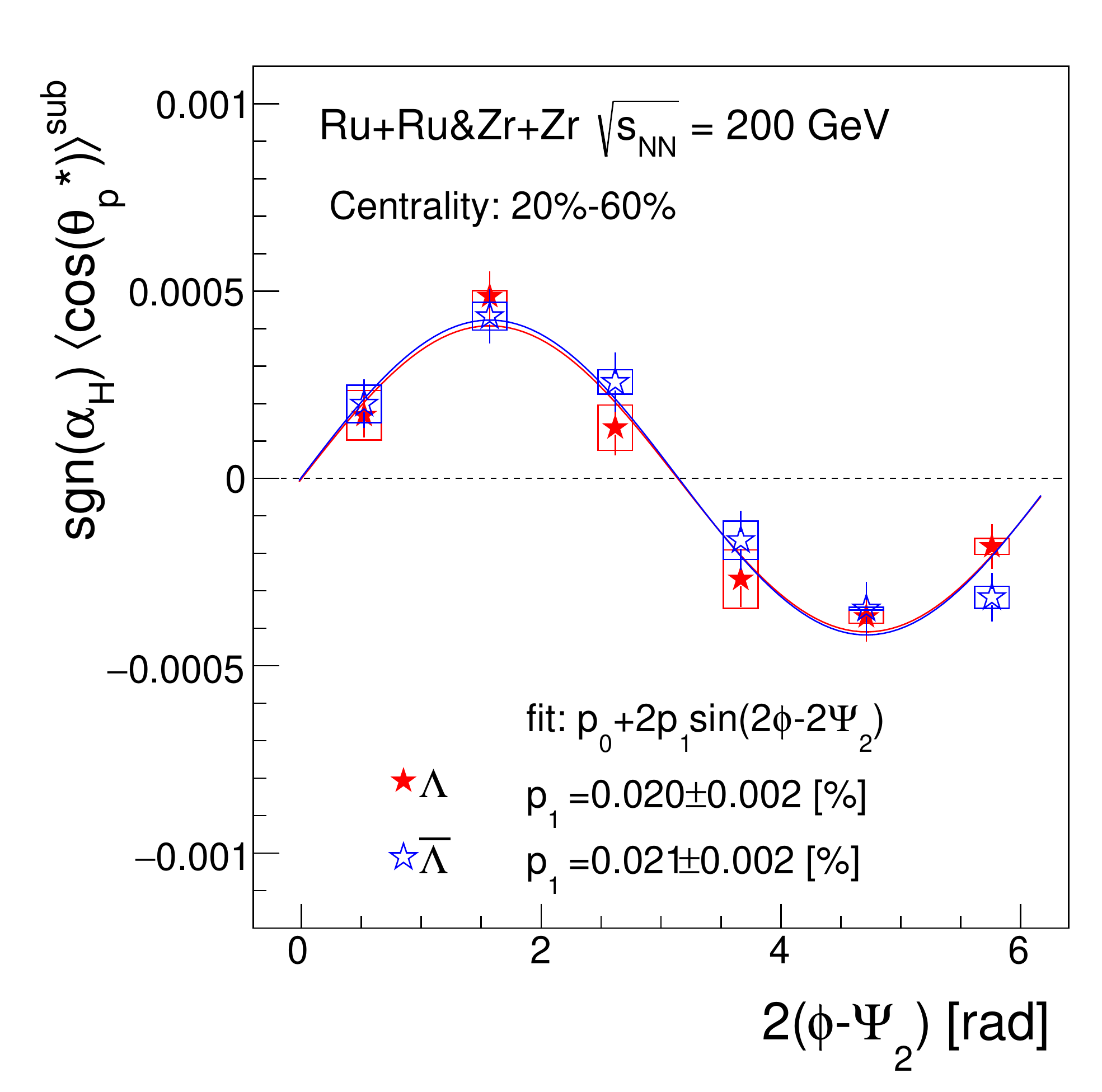}
\end{minipage}
\hspace{0.2mm}
\begin{minipage}[t]{0.49\hsize}
	\centering
	\includegraphics[width=\linewidth,clip]{Cos_vsDphi_psi2_ruzr.pdf}
\end{minipage}
\vspace{-0.3cm}
\caption{Longitudinal polarization of $\Lambda$ and $\bar{\Lambda}$ hyperons, $\langle\cos\theta_p^\ast\rangle\approx P_z\alpha_H/3$, 
as a function of azimuthal angle with respect to the second-order event plane $\Psi_2$ (left) 
and to the third order event plane $\Psi_3$ (right) in 20-60\% isobar (Ru+Ru and Zr+Zr) collisions at \snn = 200 GeV~\cite{qm2022:Niida}.
The solid lines are fit functions as indicated in the figures.}
\label{fig:Pz} 
\vspace{-0.3cm}
\end{figure}
Recent high statistics data in isobar collisions at \snn = 200 GeV taken by STAR allows to study the polarization even with higher harmonic flow plane.
Figure~\ref{fig:Pz} shows raw signal of the longitudinal polarization as a function of hyperon's azimuthal angle 
relative to the second-order (left) and third-order (right) event planes~\cite{qm2022:Niida}. Results with the second-order event plane show a clear sine modulation 
as seen in Au+Au collisions. Results with the third-order event plane also exhibit a similar sinusoidal pattern relative to the azimuthal angle, 
indicating that vorticity along the beam direction is induced by triangular flow as well as elliptic flow. 
Hydrodynamic model with the shear term~\cite{Alzhrani:2022dpi} qualitatively explain the data for both the second and third-orders
in central-midcentral collisions (but not in peripheral collisions), 
showing sensitivity of the measurement to specific shear viscosity of quark-gluon plasma formed in the collisions.

\section{Summary}
In these proceedings, recent progress on the polarization measurements has been reviewed.
Thanks to new experimental data at lower energies, energy dependence of global polarization has been studied in detail. 
Interestingly, the global polarization smoothly increases in lowering collision energy, with a possible maximum around \snn = 3 GeV,
despite the change from partonic matter to hadronic matter in studied energy range. The uncertainty is still large 
but it can be improved in future analyses or experiments. Measurements with multistrangeness are of great interest and future improvements 
will provide critical information on spin dynamics in the collisions.
New measurement on local polarization with the third-order event plane in isobar collisions indicates 
the creation of triangular-flow-driven vorticity along the beam direction, which shed light on the sign problem of the local polarization.
{~}\\

\noindent
{\bf Acknowledgment}\vspace{2pt} T. Niida is supportted by JSPS KAKENHI Grant Number JP22K03648.

%For one-column wide figures use syntax of figure~\ref{fig-1}
%\begin{figure}[h]
%% Use the relevant command for your figure-insertion program
%% to insert the figure file.
%\centering
%%\includegraphics[width=1cm,clip]{tiger}
%\label{fig-1}       % Give a unique label
%\end{figure}

%For two-column wide figures use syntax of figure~\ref{fig-2}
%\begin{figure*}
%\centering
%% Use the relevant command for your figure-insertion program
%% to insert the figure file. See example above.
%% If not, use
%\vspace*{5cm}       % Give the correct figure height in cm
%\label{fig-2}       % Give a unique label
%\end{figure*}
%
%For figure with sidecaption legend use syntax of figure
%\begin{figure}
%% Use the relevant command for your figure-insertion program
%% to insert the figure file.
%\centering
%\sidecaption
%%\includegraphics[width=5cm,clip]{tiger}
%\label{fig-3}       % Give a unique label
%\end{figure}

%For tables use syntax in table~\ref{tab-1}.
%\begin{table}
%\centering
%\label{tab-1}       % Give a unique label
%% For LaTeX tables you can use
%\begin{tabular}{lll}
%\hline
%first & second & third  \\\hline
%number & number & number \\\hline
%\end{tabular}
%% Or use
%\vspace*{5cm}  % with the correct table height
%\end{table}
%
% BibTeX or Biber users please use (the style is already called in the class, ensure that the "woc.bst" style is in your local directory)
 \bibliography{ref_hyp2022}
%\begin{thebibliography}{}
%%
%% and use \bibitem to create references.
%Journal Author, Journal \textbf{Volume}, page numbers (year)
%Book Author, \textit{Book title} (Publisher, place, year) page numbers
%\end{thebibliography}

\end{document}